# Fuel Economy Gaps Within & Across Garages: A Bivariate Random Parameters Seemingly Unrelated Regression Approach

Behram Wali
Graduate Research Assistant, Department of Civil & Environmental Engineering
The University of Tennessee
bwali@vols.utk.edu

Asad J. Khattak, Ph.D.
Beaman Distinguished Professor, Department of Civil & Environmental Engineering
The University of Tennessee
akhattak@utk.edu

David L. Greene
Research Professor, Department of Civil and Environmental Engineering &
Senior Fellow, Howard H. Baker, Jr. Center for Public Policy
The University of Tennessee
dgreen32@utk.edu

Jun Liu, Ph.D.
Travel Demand Modeler
Virginia Department of Transportation (VDOT)
Jun.Liu@VDOT.Virginia.gov

# Fuel Economy Gaps Within & Across Garages: A Bivariate Random Parameters Seemingly Unrelated Regression Approach

Behram Wali, Asad J. Khattak, David L Greene, Jun Liu

**Abstract –**The key objective of this study is to investigate the interrelationship between fuel economy gaps and to quantify the differential effects of several factors on fuel economy gaps of vehicles operated by the same garage. By using a unique fuel economy database (fueleconomy.gov), users' self-reported fuel economy estimates and government's fuel economy ratings are analyzed for more than 7000 garages across the U.S. The empirical analysis, nonetheless, is complicated owing to the presence of important methodological concerns including potential interrelationship between vehicles within the same garage and unobserved heterogeneity. To address these concerns, bivariate seemingly unrelated fixed and random parameter models are presented. With government's test cycle ratings tending to over-estimate the actual on-road fuel economy, a significant variation is observed in the fuel economy gaps for the two vehicles across garages. A wide variety of factors such as driving style, fuel economy calculation method, and several vehicle specific characteristics are considered. Drivers who drive for maximum gas mileage or drives with the traffic flow have greater on-road fuel economy relative to the government's ratings. Contrarily, volatile drivers have smaller on-road fuel economy relative to the official ratings. Compared to the previous findings, our analysis suggests that the relationship between fuel type and fuel economy gaps is complex and not unidirectional. Regarding several vehicle and manufacturer related variables, the effects do not just significantly vary in magnitude but also in the direction, underscoring the importance of accounting for within-garage correlation and unobserved heterogeneity for making reliable inferences.

## INTRODUCTION

In 2016, more than 143 billion gallons of finished motor gasoline was consumed in the United States which is the highest on record (EIA, 2017). This translates to a daily average of approximately 392 million gallons of gasoline consumption. As a result, the oil dependency of the United States is linked with significant external costs, largely related to geopolitical, environmental, and economic concerns (Jacobsen, 2013). One possible way to reduce oil dependency is to improve fuel economy, which is a concern for consumers, vehicle manufacturers, and policy makers (Greene et al., 2017 ). The U.S. Department of Energy is required to provide consumers with fuel economy (miles per gallon) ratings for passenger cars and light-duty trucks (EPA, 2015; EIA, 2016). Accurate and unbiased fuel economy estimates can help governments to enforce and achieve set targets for CAFE and greenhouse gas (GHG) emissions (EIA, 2016), enabling auto manufacturers to increase sales by cost-effective fuel economy improvements (An and Sauer, 2004; Anderson et al., 2011), and assist consumers in making informed comparisons among vehicles in vehicle purchase decisions (Helfand and Wolverton, 2009; Liu et al., 2016b; Greene et al., 2017b).

Specifically, two sets of fuel economy estimates, test-cycle and label, are provided for every make, model, engine and transmission configuration (EPA, 2015). To evaluate automakers' compliance with government regulations, the test-cycle estimates are produced under controlled laboratory conditions where vehicles are tested over city and highway driving cycles. On the other hand, the label estimates build upon the test-cycle estimates by adding three additional cycles for aggressive driving, city driving in colder temperatures, and air conditioning use. However, the extent to which the government's fuel economy ratings is useful for vehicle comparison purposes is extensively debated (McNutt et al., 1982; ConsumerReports, 2013; Dings, 2013; Mock et al., 2013). Albeit less biased, the fuel economy estimates cannot accurately predict the real-world fuel economy for a given motorist, with studies suggesting that government ratings tend to over-estimate on-road fuel economy by 15-20% (McNutt et al., 1978; McNutt et al., 1982; Rykowski et al., 2005; Lin and Greene, 2011; Huo et al., 2012; Dings, 2013; Mock et al., 2013; Greene et al., 2017b).

Several important factors such as vehicle type, fuel type, driving behavior, driving conditions, and methods of estimating on-road fuel economy are found associated with the gap between on-road and EPA fuel economy ratings (McNutt et al., 1982; Mintz et al., 1993; Berry, 2010; Lin and Greene, 2011). However, little is known about the degree of correlation between fuel economy gaps of vehicles within the same garage. Fuel economy gap can be defined as the differences between Environmental Protection Agency (EPA) fuel economy ratings and actual on-road fuel economy of road users (Greene et al., 2017a). Such an analysis is important for two main reasons. First, an understanding of the correlation between the fuel economy gaps in the same garage can shed light on the usefulness of government's ratings for vehicle comparison purposes (Greene et al., 2017b). To be useful for relative comparisons a consumer who experiences lower fuel economy than the official estimates driving one vehicle should experience similarly lower fuel economy driving other vehicles. In other words, deviations of real world fuel economy from the official ratings should be correlated across vehicles for the same driver. Second, fuel economy gaps of vehicles operated by same garages are interrelated and all the factors known to affect the fuel economy gaps may not be available. Having said this, a simultaneous empirical analysis is warranted to fully capture the interdependencies and better understand the factors associated with fuel economy gaps within the same garage.

Thus, the objective of the current study is to conceptualize and quantify the differential effects of factors associated with fuel economy gaps of vehicles in the same garage. To achieve this, an in-depth statistical analysis is conducted of users' self-reported fuel economy data vis-a-vis EPA fuel economy ratings for more than 7000 garages who voluntarily submitted data to the "My MPG" section of the government website www.fueleconomy.gov (Greene et al., 2017b; Greene et al., 2015). The scope of this study is limited to the My MPG sample where people entered data on two vehicles in their My MPG. It is important to note that these are not necessarily two-vehicle garages. In addition, as the data on two vehicles are entered by the same person, it is highly likely that the two-vehicles belong to the same garage. However, given the possibility that the two vehicles may yet be operated by different garages, we use the term "garage" from this point onwards. The empirical analysis, nonetheless, is complicated owing to the presence of important methodological concerns including potential interrelationships between fuel economy gaps of vehicles in the same garage and unobserved heterogeneity. Given these significant concerns (discussed later in detail), a joint model estimation framework is presented that allows for consideration of within-garage correlation and for the effects of explanatory factors to vary both for vehicles and across different garages in a multivariate empirical setting.

## LITERATURE REVIEW

### Fuel Economy Gap

The official government fuel economy ratings (both test cycle and label) are based on pre-defined standard driving cycles in controlled laboratory experiments. Real-world driving is a complex task and all the factors that may determine on-road fuel economy cannot be controlled in laboratory settings. Thus, pre-defined driving cycles cannot precisely represent the driving behavior of individual drivers. Given that precise fuel economy estimates are crucial to the usefulness of EPA fuel economy estimates, several researchers have focused on investigating the gaps between actual on-road fuel economy and the EPA estimates. Collectively, researchers have found that the EPA ratings tend to over-estimate on-road fuel economy by 15-20% (McNutt et al., 1978; McNutt et al., 1982; Rykowski et al., 2005; Lin and Greene, 2011; Huo et al., 2012; Dings, 2013; Mock et al., 2013; Greene et al., 2017b).

An empirical analysis of over 3,000 self-reported individual drivers' fuel economy estimates concluded that the EPA estimates may be unbiased predictors of the "average" fuel economy that road users may get (Greene et al., 2006a). While unbiased, Greene et al. (2006) found that the EPA estimates are highly imprecise predictors of individual drivers' fuel economy with a wide 95% confidence interval of ± 7 mpg (Greene et al., 2006a). Likewise, rigorous statistical analysis of a much larger sample (N = 67000) of individual fuel economy estimates concluded a two-standard deviation interval of ± 10 mpg (around mean label of 24 mpg) for gasoline vehicles (Greene et al., 2017b). Besides the academia and research community, the topic of fuel economy gap is also of interest to public and media. For example, a study by Consumer Reports in 2005 concluded a 9% and 18% gap (difference between on-road fuel economy and EPA label estimates) for conventional gasoline and hybrid vehicles respectively (ConsumerReports, 2013). A 2013 study, however, revealed that the gap has decreased to 2% and 10% for gasoline and hybrid vehicles respectively (ConsumerReports, 2013). However, the extant literature also possesses some evidence that the gaps between on-road fuel economy and test cycle estimates have been increasing since 2005 (Greene et al., 2017b). Previous studies raise important questions about the use of official fuel economy estimates as a predictor of on-road fuel economy. Nonetheless, no direct evidence is presented about the usefulness of official ratings for making comparisons among vehicles. That is, broadly speaking, if a garage experiences a 10% gap on their first vehicle, can the garage expect a similar gap on their second vehicle too?

**Correlates of Fuel Economy**

Researchers have, for decades, investigated factors that may be associated with fuel economy. Several factors such as vehicle type, fuel type, driving behavior, driving conditions, and methods of estimating on-road fuel economy are found associated with the fuel economy itself, or the gaps between on-road and the EPA fuel economy ratings (McNutt et al., 1982; Mintz et al., 1993; Greene et al., 2006a; Sharer et al., 2007; G, 2008; Berry, 2010; Lin and Greene, 2011; Greene et al., 2017b). Compared to front-wheel drive, manual transmission, and passenger cars, rear-wheel drive, automatic transmission, and trucks are reported to have a higher gap (McNutt et al., 1982; Mintz et al., 1993; Greene et al., 2017b). Likewise, hybrid vehicles are found to exhibit higher gaps and variability compared to conventional vehicles (Sharer et al., 2007). The correlations between driving behavior and fuel consumption have also been widely studied (Ma et al., 2015) (Kurani et al., 2015). Driving at optimal speeds under free-flow traffic conditions and eco-driving are both reported correlated with lower fuel consumption (Kurani et al., 2015). Another recent study by Ma et al. (2015) developed a real-world data-driven vehicle-engine combined model and proposed 26 different driving style parameters relevant to fuel consumption of city buses under different driving regimes, such as accelerating, normal running, and decelerating processes (Ma et al., 2015). Importantly, the study concluded that driving style characteristics in acceleration process are critical for fuel consumption of city buses under varying conditions (Ma et al., 2015). Regarding effects of driving behavior on fuel economy gap, a recent study found that drivers who drive for achieving maximum fuel economy or drive cautiously tend to reduce their on-road gap (Greene et al., 2017b). Numerous other factors such as vehicle technology, weather, road grade, and vehicle age are also found correlated with on-road fuel economy (Huo et al., 2012; Tang et al., 2015).

***Research Gap and Methodological Issues:***

The above studies did not focus on fuel economy gaps of vehicles within the same garage. Given several observed and unobserved factors common to each vehicle within a garage, an analysis of vehicle fuel economy gaps within a garage is warranted and how they vary within and across garages. As acknowledged in recent literature (Greene et al., 2006b; Greene et al., 2017 ), an understanding of the correlation between the fuel economy gaps in the same garage can shed light on the usefulness of government's ratings for vehicle comparison purposes (Greene et al., 2006b; Greene et al., 2017 ). Next, regarding the factors correlated with fuel economy gap, previous studies did not account for the interdependencies of fuel economy gaps within a garage and unobserved heterogeneity ignoring which can lead to biased inferences and may hide important information embedded in data.

The present study focuses on investigating fuel economy gaps and associated factors within a garage. To achieve this, an in-depth statistical analysis is conducted on users' self-reported fuel economy data vis-a-vis EPA fuel economy ratings for more than 7000 two-vehicle garages. A unique publicly available database is used which provides fuel economy data by make, model, engine, transmission, and the state in which vehicles are driven. The significant methodological concerns are briefly discussed next.

*Within-garage Correlation and Unobserved Heterogeneity:*

Data relating to all the factors known to affect the fuel economy gaps of the two vehicles within a garage may not be available. While not known to the analyst, such factors can create interrelationships between the fuel economy gaps of vehicles, and thus warrant a joint model estimation framework. Related to the observed and unobserved factors that may determine fuel economy gaps, and within the joint estimation framework, a second important methodological concern relates to unobserved heterogeneity which is reflective of the possibility of varying parameter estimates both for vehicles within a garage and across different garages. As the effects of several factors on fuel economy gaps are highly context (garage) specific, ignoring unobserved heterogeneity can lead to erroneous inferences, biased, inefficient and/or inconsistent parameter estimates (Mannering et al., 2016).

*Self-Selection and Omitted Variable Bias:*

The proposed methodology also attempts to address important concerns that primarily arise due to the self-selected and non-random nature of the fueleconomy.gov's My MPG sample. For example, an argument can be made that users who participated in fueleconomy.gov's My MPG project have a higher tendency to exceed their EPA test cycle estimates than those who did not participate in the My MPG project. From a conceptual standpoint, this may display an omitted variables specification problem (Heckman, 1977; Heckman, 2010), that is, a systematic bias arising from uncontrolled or unobserved differences in fuel economy gaps between those who opted to participate and those who did not. Applying simple linear regression models, in this case, will generate a systematic bias mainly characterized as a simple specification error or omitted variable problem (Khattak and Rodriguez, 2005; Mannering et al., 2016), and which can result in biased and inconsistent parameter estimates. Given the reality that the My MPG sample is non-random and self-selected, obtaining data on all factors that may have affected the fuel economy gaps is not possible. As such, to mitigate the adverse impacts of omitting significant explanatory factors, we consider a methodological remedy by allowing heterogeneity in the effects of explanatory variables through a random-parameter modeling specification (Mannering et al., 2016).

Having said this, the present study empirically contributes by addressing the above methodological concerns in a multivariate context. The proposed framework seeks to provide new insights on interrelated fuel economy gaps within a garage. Specifically, bivariate seemingly unrelated fixed- and random-parameter modeling approaches are formulated that account for the presence of correlation within a garage and unobserved heterogeneity[1]. To authors' knowledge, such empirical methods have not been used in the context under discussion.

[1] The use of bivariate/multivariate analysis differs across fields. In some fields, bivariate analysis is used for analysis of one dependent variable and one independent variable. Whereas, in other fields, bivariate (or multivariate) analysis refers to simultaneous modeling of two or more than two dependent variables (Weiss, 1993; Yamamoto and Shankar, 2004b; Greene et al., 2008; Anastasopoulos et al., 2012a; Russo et al., 2014; Wali et al., 2017b; Wali et al., 2017c). In this paper, we mean to refer to the latter case where two dependent variables (fuel economy gaps of two vehicles) are modeled simultaneously.

# METHODOLOGY

## Empirical Setting

The data used in this study is obtained from the "My MPG" section of the government website www.fueleconomy.gov, jointly managed by the U.S. Department of Energy and the Environmental Protection Agency (EPA). The website allows users to share their on-road fuel economy estimates for vehicles of given make, model year, transmission, and engine. Importantly, the website also provides users the option to compare their fuel economy estimates with the official EPA test cycle estimates (Greene et al., 2006a). Since 2004, a total of 77,126 entries have been made by users, with observations from every state and for every model year of vehicle from 1985 to 2014. The website has several internal tools to check for plausibility and errors (Greene et al., 2017 ). For detailed discussion about the data quality, representativeness, and error checking tools, interested readers are referred to (Greene et al., 2006a; Lin and Greene, 2011; Greene et al., 2017b; Greene et al., 2017 ; Greene et al., 2015).

Given the objectives of the present study, a sub-sample of garages with data on two vehicles is extracted from the original 77,126 entries. By using anonymous "DriverID" which represents a unique vehicle, garages with two vehicles are identified (represented by GarrageID on the website). Based on model year, the older vehicle is labelled as vehicle 1 and the newer vehicle labelled as vehicle 2. Next, for each of the two vehicles in a garage, relevant data including transmission, wheel drive, vehicle make, model, year, body type, and manufacturer are extracted and linked. For each of the two vehicles, information about the method used for calculating the on-road fuel economy, driving style, and state in which the vehicle is driven are also available and used. The large number of observations and detailed information about the on-road fuel economy with respect to vehicle characteristics make this a unique and rich publicly available source of information. Given that the fuel economy estimates are self-reported, the data may be representation of what people "think" they get rather than "what they actually get." Our implicit assumption is that perception measures/relates to reality. However, note that our methodological approach accounts for varying levels of preference heterogeneity (as discussed earlier).

To quantify the gap for each vehicle in the garage, the key variables used are the user reported My MPG and the EPA test cycle ratings for each of the two vehicles. The gap is thus quantified as a ratio of user-reported on-road MPG and EPA test cycle estimates. This measure allows for intuitive comparison of gap within a garage and across garages. Based on the preliminary analysis of the bivariate distribution of fuel economy gap of vehicle 1 and 2, some obvious outliers were spotted. As such, formal outlier analysis was conducted by identifying fuel economy gaps for the two vehicles that were outside the $\mu \pm 2\ standard\ deviations$ and $\mu \pm 3\ standard\ deviations$ intervals. A total of 270 and 233 observations were found to be outside the $\mu \pm 2SD$ interval for vehicle 1 and 2 respectively, whereas only 69 and 49 observations were found to be outside the $\mu \pm 3SD$ interval respectively. As a result, a total of 118 observations were deleted based on the $\mu \pm 3SD$ analysis. Narrower threshold i.e., $\mu \pm 2SD$, was not used as it created obvious linear boundaries in the bivariate distribution of the two vehicles' fuel economy gaps. Note that, removing the outliers based on $\mu \pm 3SD$ did not significantly alter the resulting distributions, and are in fact very similar to the untrimmed sample (discussed later).

## Statistical Methods

The current study focuses on identifying and better understanding factors that may be associated with fuel economy gaps of two vehicles within a garage. Thus, statistical models are developed for two variables: 1) fuel economy gap (ratio of user reported My MPG and EPA ratings) of vehicle 1, and 2) fuel economy gap of vehicle 2. Given the continuous nature of the two response outcomes, linear regression/ordinary least squares regression (OLS) is an obvious choice (Washington et al., 2010). The model formulation is:

$$SF_{veh-1,i} = \boldsymbol{\beta}_{veh-1}\boldsymbol{X_i} + \xi_{veh-1,i} \quad (1)$$
$$SF_{veh-2,i} = \boldsymbol{\beta}_{veh-2}\boldsymbol{X_i} + \xi_{veh-2,i} \quad (2)$$

Where: $SF_{veh-1,i}$ and $SF_{veh-2,i}$ are the fuel economy gaps for vehicle 1 and 2 in garage $i$ respectively; $\boldsymbol{X_i}$ is a vector of set of exogenous explanatory variables associated with fuel economy gaps $SF_{veh-1,i}$ and $SF_{veh-2,i}$; $\boldsymbol{\beta}_{veh-1}$ and $\boldsymbol{\beta}_{veh-2}$ are the column vectors of estimable parameters corresponding to explanatory factors in Equation 1 and 2; and $\xi_{veh-1,i}$ and $\xi_{veh-2,i}$ are error terms corresponding to fuel economy gaps of vehicle 1 and 2 respectively. Efficient and unbiased parameter estimates can be obtained if the standard assumption is fulfilled, i.e., information on all factors that may affect fuel economy gaps of the two vehicles is available. Otherwise, the efficiency of parameter estimates is likely to be compromised (Washington et al., 2010).

Among many sources of missing information, one is the potential possible correlation between the error terms of fuel economy gaps of vehicle 1 and 2. There are several reasons to expect the correlation in error terms[2,3]. Data relating to many factors known to affect the fuel economy gaps of the two vehicles within a garage may not be available, and which may result in correlation between the fuel economy gaps of the two vehicles. If this contemporaneous correlation (generated by different unobserved factors) is accounted for, more efficient parameter estimates can be obtained. Thus, the conventional seemingly unrelated regression equation (SURE) estimation framework can be formulated as (Zellner, 1962; Anwaar et al., 2012):

$$SF_{veh-1,i} = \boldsymbol{\beta}_{veh-1}\boldsymbol{X_i} + \boldsymbol{\gamma}_{veh-1}\boldsymbol{Z_i} + \xi_{veh-1,i} \quad (3)$$
$$SF_{veh-2,i} = \boldsymbol{\beta}_{veh-2}\boldsymbol{X_i} + \boldsymbol{\gamma}_{veh-2}\boldsymbol{Z_i} + \xi_{veh-2,i} \quad (4)$$

Where: $\boldsymbol{Z_i}$ is a vector of characteristics common to both equations**;** $\boldsymbol{X_i}$ is a vector of characteristics unique to both equations; $\beta$ and $\gamma$ are the vectors of estimable parameters, and the error terms $\xi_{veh-1,i}$ and $\xi_{veh-2,i}$ are now assumed to follow a bivariate normal distribution as with mean 0 and variance $\sigma^2$, correlation $\rho$, and covariance matrix variance-covariance matrix $\sum$. Altogether, the formulation becomes:

$$\begin{pmatrix}\xi_{veh-1}\\ \xi_{veh-2}\end{pmatrix} \sim N\left[\begin{pmatrix}0\\0\end{pmatrix}, \begin{pmatrix}\sigma^2_{\xi_{veh-1}} & \sigma_{\xi_{veh-1}}\sigma_{\xi_{veh-2}}\\ \sigma_{\xi_{veh-2}}\sigma_{\xi_{veh-1}} & \sigma^2_{\xi_{veh-2}}\end{pmatrix}\right] \quad (6)$$

Where: $\sigma^2_{\xi_{veh-1}}$ is the variance associated with $SF_{veh-1}$, $\sigma^2_{\xi_{veh-2}}$ is the variance associated with $SF_{veh-2}$, and $\sigma_{\xi_{veh-1}}\sigma_{\xi_{veh-2}}$ is the covariance between $SF_{veh-1}$ and $SF_{veh-2}$. Once the variance-covariance matrix is estimated for fixed- and random-random parameter models, the correlation coefficient $\rho$ can be computed as $\rho = \frac{\sigma_{\xi_{veh-1}}\sigma_{\xi_{veh-2}}}{\sigma_{\xi_{veh-1}}\sigma_{\xi_{veh-2}}}$. Several methods such as generalized least squares (GLS) or maximum likelihood estimation can be used to estimate the system of equations in Equation 3 and 4 (Zellner, 1962) (Anwaar et al., 2012). Under GLS, the system is solved by initiating with OLS estimates as obtained in Equation 1 and 2 above:

$$\widehat{\boldsymbol{\beta}}_{veh-i} = (\boldsymbol{X}^T\boldsymbol{X})^{-1}\boldsymbol{X}^T\boldsymbol{SF}_{veh-i} \tag{7}$$

[2] Note the at the equation systems (vehicle 1 and 2) shown in Equation 1 does not include direct interactions among the dependent variables, i.e., $SF_{veh-1,i}$ is not an independent variable in the equation for $SF_{veh-2,i}$, and vice versa. This means that the fuel economy gap of vehicle 1 does not directly determine the fuel economy gap of vehicle 2. However, are inter-related because different garage factors are common to both vehicles, and these common unobserved factors can create a dependency in $\xi_{veh-1,i}$ and $\xi_{veh-2,i}$.

[3] As the two vehicles belong to a same garage/household, different unobserved characteristics can be common to both vehicles and this may generate correlation between the fuel economy gaps of the two vehicles. As an example, for this dataset, such unobserved factors can include (to name a few) household income, household overall education status, vehicle weight, and several environmental and weather related factors. On the other hand, the fuel economy gaps of the two vehicles within a household can also be interrelated due to several observed factors. Such observed factors can include vehicle specific factors such as vehicle body type or vehicle class, fuel type, and geographic region, among many others. These common observed and unobserved factors are of interest in the sense that it poses a methodological challenge of potential correlation between fuel economy gaps of the two vehicles under consideration, as discussed in detail elsewhere in the context of transportation safety (Yamamoto and Shankar, 2004a; Washington et al., 2010; Mannering and Bhat, 2014; Sarwar and Anastasopoulos, 2017; Wali et al., 2017b; Wali et al., 2017c).

Where: $\widehat{\boldsymbol{\beta}}_{veh-i}$ is the $\tau \times 1$ column vector (with $\tau$ number of estimable parameters), $\boldsymbol{X}$ is an $N \times \tau$ matrix of data on exogenous explanatory factors (N is sample size), $\boldsymbol{X}^T$ being transpose of $\boldsymbol{X}$, and $\boldsymbol{SF}_{veh-i}$ is the column vector of response outcomes (fuel economy gap for vehicle $i$). To account for correlation among the residuals, Equation 5 is re-written by inserting $\Omega$ matrix (Washington et al., 2010), and $\Omega$ matrix is estimated using individual OLS estimates (as in Equation 5) to finally arrive at parameter estimates of SURE model as (Washington et al., 2010; Anwaar et al., 2012):

$$\widehat{\boldsymbol{\beta}}_{veh-i} = (\boldsymbol{X}^T\boldsymbol{\Omega}\boldsymbol{X})^{-1}\boldsymbol{X}^T\boldsymbol{\Omega}\boldsymbol{SF}_{veh-i} \tag{8}$$

The efficiency of estimates obtained in Eq. 6 will increase as $\rho$ increases. The traditional SURE regression framework does not account for unobserved heterogeneity in that a single parameter is estimated for all the observations in the sample (Mannering et al., 2016). By estimating a single parameter estimate, the standard SURE model will imply that the effect of real-wheel drive is positive (as an example) on the fuel economy gaps of vehicle 1 and 2, and this may be a very restrictive and unrealistic assumption. Due to several observed and unobserved factors, there is a real possibility that the effects of explanatory factors on both outcomes may vary across garages. Also, in order to establish reliable correlations between observed factors and fuel economy gaps, it is crucial to account for the effects of unobserved factors (i.e., omitted variable bias) (Greene et al., 2006b; Mannering et al., 2016).

Having said this, we introduce random parameters in the joint systems of equation framework by estimating[4]:

[4] Here, we distinguish between random parameter modeling techniques, and traditional fixed-and random-effect methods. In traditional econometrics, fixed-effects and random-effects modeling approaches are two typical techniques to incorporate individual-specific unobserved heterogeneity in a "panel data" setting (Greene and Hensher, 2010; Wooldridge, 2010). That is, before proceeding to random parameter models, a panel or group data setting is usually employed, fixed- and random-effects estimators used, and then proceed to random parameter/random coefficients models in panel or group data (Mannering et al., 2016). In particular, with fixed or random effects estimators, the individual-specific unobserved heterogeneity is captured in the constant term/intercept in a panel data setting (Greene and Hensher, 2010; Wooldridge, 2010). As discussed in detail in Wooldridge (2010), the unobserved individual heterogeneity in a random-effects framework is assumed to be unrelated to the observed independent variables, an assumption difficult to justify in practice (Wooldridge, 2010). While this assumption can be relaxed in a fixed-effects approach, the significant problem of incidental parameters in fixed-effects estimation remain (please see page 60 of Greene and Hensher (2010) and (Mannering et al., 2016)). Importantly, while fixed- and random-effects methods typically necessitate panel data approach, random parameter modeling techniques do not, i.e., random parameter models can be estimated in a cross-sectional setting too as is the case in the present study (Train, 2003; Mannering et al., 2016). Compared to fixed- and random-effects estimators, the random parameter modeling technique allows the possibility of coefficients on other explanatory variables to vary across individual observations (in case of cross-sectional data) and groups (in case of panel data) (Chapter 11 of (Greene, 2011)). As such, the random parameter modeling technique represents a very useful extension in which analysts can broaden the amount of unobserved heterogeneity across individual observations, garages/households in our case. For further details, see (Greene and Hensher, 2010).

$$SF_{veh-1,i} = \boldsymbol{\beta}_{\boldsymbol{i}_{veh-1}} \boldsymbol{X_i} + \boldsymbol{\delta}_{veh-1} \boldsymbol{B_i} + \xi_{veh-1,i} \quad (9)$$
$$SF_{veh-2,i} = \boldsymbol{\beta}_{\boldsymbol{i}_{veh-2}} \boldsymbol{X_i} + \boldsymbol{\delta}_{veh-2} \boldsymbol{B_i} + \xi_{veh-2,i} \quad (10)$$

Where: $\boldsymbol{\beta_i}$'s are vehicle and garage specific parameter estimates for random parameters, $\boldsymbol{\delta}$ are the vehicle specific parameter estimates for fixed parameters[5]. The individual garage and vehicle specific parameter estimates can be written as:

$$\boldsymbol{\beta}_{\boldsymbol{i}_{veh-1}} = \boldsymbol{\beta}_{veh-1} + \boldsymbol{\zeta_i} \quad (11)$$
$$\boldsymbol{\beta}_{\boldsymbol{i}_{veh-2}} = \boldsymbol{\beta}_{veh-2} + \boldsymbol{\varsigma_i} \quad (12)$$

Where: $\boldsymbol{\beta}_{veh-1}$ and $\boldsymbol{\beta}_{veh-2}$ are the mean vehicle specific regression coefficients; $\boldsymbol{\zeta_i}$ and $\boldsymbol{\varsigma_i}$ are the randomly distributed deviations over the sample of two-vehicle garages that has an assumed distribution[6]. Several distributions are possible for the random parameters such as normal, log-normal, logistic, Weibull, triangular, and Erlang distributions. The estimation of random parameter SURE model requires integration of the likelihood over unobserved random effects/random parameters (Train, 2003). Thus, we use a maximum simulated likelihood technique where specific number of draws are taken per observation (and for each equation) with the sequence of random draws as Halton sequence (Train, 2003). The extant literature suggests that a Halton sequence of 200 draws can provide accurate parameter estimation with significant parameter stability (Train, 2003). For this application, we have used 400 Halton draws for integrating the likelihood over several unobserved random parameters[7].

*Models Comparison:*

Finally, to evaluate the goodness-of-fit performance of competing fixed- and random-parameter SURE models, we score each model by different information criteria. Specifically, Akaike's Information Criteria (AIC), Bozdogan's dimension consistent AIC (CAIC), Schwartz Bayesian Information Criteria (BIC), and Bozdogan's criteria of information complexity (ICOMP) are used that are reported well-suited to compare nested and non-nested models (Bozdogan, 2000). These criteria are used to select the "best" model compromising an adequate goodness of fit and a small number of parameters by adding a penalty for overparameterization to the lack of fit measure (estimate of the maximum likelihood or the residual sum of squares). The best model minimizes the criterion. The three informational measures differ in the penalty term, SBIC penalizing more severely for larger samples, and ICOMP accounting for covariance structure of a model (and, thus, for collinearity between the factors and dependence among the parameter estimates). Following (Bozdogan, 2000), the different goodness-of-fit criteria can be scored as:

---

[5] We remind that the possibility of both fixed and random parameters in a same model specification should not be confused with "fixed-effect" or "random-effect" models where either a fixed-effect or random effect can be incorporated in a same model specification. In random parameter modeling framework (which we present in this paper), some explanatory variables can be treated as fixed parameters whereas others can be treated as random parameters. The reason is that the random parameter terminology refers to the possibility of varying parameter estimates across individual observations. In our context, it means that a random parameter refers to an explanatory factor for which the parameter estimates ($\beta$) are allowed to vary across observations as opposed to a fixed/constant $\beta$. As a result, based on empirical testing (as discussed in detail in the paper), some explanatory variables can be treated as random parameters whereas other can be fixed parameters in a same model specification.

[6] Given the nature of the fuel economy data, note that the seemingly unrelated random parameter modeling framework presented in this study is applied in a purely cross-sectional data setup. As such, the random parameters are individual observations specific where each observation refers to an individual household, and as such are not group specific. Nonetheless, the random parameter modeling framework is readily extendable to panel data setup (Venkataraman et al., 2013). Further details can be found in (Anastasopoulos, 2016; Behnood and Mannering, 2017; Li et al., 2017b; Wali et al., 2017a; Wali et al., 2018).

[7] Given the large number of potential explanatory factors, a series of joint models with random parameter specification for different sets of explanatory variables were estimated and compared formally through model selection techniques. Despite the simulation based estimation used in this study, which is generally faster and more efficient than sparse-grid adaptive quadrature (Heiss and Winschel, 2008), each model run took 7-8 hours to converge on a work-station level computer (Dell Precision T7600, 3.1 GHZ (32 CPUs)).

$$\mathrm{AIC} = -2\ln L(\theta'_k) + 2k \quad (13)$$
$$\mathrm{CAIC} = -2\ln L(\theta'_k) + k[\ln(n) + 1] \quad (14)$$
$$\mathrm{SBIC} = -2\ln L(\theta'_k) + k\ln(n) \quad (15)$$
$$\mathrm{ICOMP} - \mathrm{IFIM} = -2\ln L(\theta'_k) + s\ln tr\left(\frac{I^{-1}(\theta'_k)}{s}\right) - \ln|I^{-1}\theta'_k)| \quad (16)$$

Where: $n$ is the sample size, $k$ is the number of estimable parameters in the model, $\theta'$ are the estimated coefficients, and $I^{-1}\theta'_k$ is specifically the inverse of Fisher information matrix, as $I^{-1}\theta'_k = \widehat{Var}(\theta'_k)$, and finally $s = krI^{-1}\theta'_k$ (Andrews and Currim, 2003). The best model will be the one with lowest scores of afore-mentioned information criteria.

## RESULTS

### Dependent Variables

The distributions of the trimmed (excluding outliers) and untrimmed (including outliers) dependent variables, i.e., fuel economy gap of vehicle 1 and 2, are presented in the top panel of Table 1. In particular, the fuel economy gaps are calculated by dividing user reported My MPG on test cycle EPA ratings (Table 1), and which provides an intuitive measure to quantify the gaps between user reported MPG and EPA ratings. The descriptive statistics (in terms of means and standard deviations) of the trimmed sample (including outliers) and untrimmed sample (excluding outliers) are approximately similar (Table 1). For the trimmed and untrimmed sample, the user reported on-road fuel economies for vehicle 1 are 86% and 87% of the EPA test cycle ratings respectively (Table 1). Likewise, for vehicle 2, the user reported My MPG on average is 85% and 86% of the EPA test cycle ratings for the trimmed and untrimmed sample respectively (Table 1). Importantly, keeping in view the mean values, the standard deviations of the fuel economy gap of vehicle 1 and vehicle 2 suggest significant variations across the sampled garages (Table 1). Regarding absolute differences, EPA test cycle ratings tend to over-estimate the actual on-road fuel economy, i.e., EPA test cycle ratings are onaverage 3.81 miles and 4.90 miles more than the user reported on-road fuel economy, for vehicle 1 and 2 respectively (Table 1). The “My MPG” database also includes information about model year for each of the two vehicles in a garage. Thus, the classification of vehicle 1 and vehicle 2 is not arbitrary, and is designated based on model years, i.e., the older vehicle is considered vehicle 1 and newer vehicle considered vehicle 2. As such, 37.8% of “vehicle 1” in all the 7126 garages are manufactured between 1984-1998, compared to only 10.4% of “vehicle 2” in the same group (Table 1). Almost half of “vehicle 2” are produced between 2004-2008 compared to only 25% of “vehicle 1” manufactured between 2004-2008.

### Variations in Fuel Economy Gap

In order to spot differences in the fuel economy gaps of the two vehicles within a garage by US divisions and model years, the top and bottom panels of Figure 1 illustrate the calendar heat map matrices for vehicle 1 and 2 respectively. The rows of the matrices indicate the nine US divisions where the columns indicate model years. Due to space constraints, the definitions of nine US divisions are not provided here and can be found at https://en.wikipedia.org/wiki/List_of_regions_of_the_United_States. As expected and for reasons mentioned earlier, the fuel economy gap has been consistently below 1.0 (i.e., on-road MPG is less than EPA test cycle estimates) for most of the US divisions and model years (Figure 1). Importantly, for all the nine US divisions, the fuel economy gaps are increasing for the newer model year vehicles (Figure 1). The gap between user on-road MPG and EPA test cycle estimates for the two vehicles is increasing since 2005 with significant gap during 2009-2014 (refer to the Figure 1 and the density bars). For vehicle 1, and across all the model years, the average gap is 13.1% with the highest gap of 18.7% for vehicles 1 with model years between 2009-2014. Likewise, for vehicle 2, the average gap is 14.8% with the highest gap of 19.3% for vehicles 2 with model years between 2009-2014. These findings are in agreement with previous studies which documented some evidence of increasing gap for newer vehicles (Mock et al., 2013; Greene et al., 2017b), however previous studies considered vehicles independently and not within garages.

**Bivariate Distributions & Descriptive Statistics**

As the study focuses on simultaneous modeling of fuel economy gaps in two-vehicle garages, Figure 2 presents the bivariate distribution and the marginal distributions. It can be observed that the fuel economy gaps of two vehicles within the same garage are mildly positively correlated with a Pearson correlation of 0.40 (Figure 2). The density of the bivariate correlation is higher in the middle of the distribution (fuel economy gaps between 0.75 and 0.9) with smaller density elsewhere (Figure 2). This suggests that garages with fuel economy gaps in between 0.75 and 0.9 for the first vehicle are also likely (albeit with seemingly mild correlation) to experience similar fuel economy gaps for the second vehicle. From a policy perspective, the observed correlation between the fuel economy gaps seems not strong enough (correlation of 0.40) to conclude that a garage who experiences a 10% gap on one vehicle may expect a similar gap on their second vehicle too. A perfect linear correlation between the fuel economy gaps of the two vehicles will suggest that all points would lie on the y = x diagonal line (Figure 2). Expecting all points to lie exactly on the diagonal line in real-world may be unrealistic; nonetheless, the bivariate pair of points should be tightly clustered along the diagonal line for the fuel economy estimates to be greatly useful for vehicle comparison purposes (Figure 2). As can be seen in Figure 2, the points tend to be closer to the y = x diagonal line for a very small range (0.75-0.90) and are highly spread out for the rest of the range. Nonetheless, from an empirical standpoint, it is important to note that a mild correlation does not rule out the benefits of joint estimation framework. A broad spectrum of studies have shown that accounting even for the weak or mild degree of correlation in joint estimation framework can result in significant and compelling efficiency gains[8] (Khattak et al., 1995; Yamamoto and Shankar, 2004a; Anastasopoulos et al., 2012b; Wali et al., 2017b).

**(PUT TABLE 1 ABOUT HERE)**

**Descriptive Statistics**

The government's www.fueleconomy.gov database provides information about several factors that may be related to fuel economies such as driving style, methods by which users calculated their fuel economy, fuel type, transmission, wheel drive, vehicle class, and manufacturer (Table 1). The descriptive statistics for both vehicles within the garage are provided in Table 1. The information supplied by many participants about method of estimation and driving style enables additional inferences. For instance, 11.7% and 6.5% of the vehicle 1 drivers reported themselves as "drivers who drives a little faster than most" and "Accelerates quickly and passes" respectively. Given that the My MPG sample consists of self-reported entries, one may expect that users who monitor their fuel economy and submit data to www.fueleconomy.gov website may be more likely eco-drivers. Surprisingly, only 0.7% and 5.4% of the drivers described themselves as ones who "drives for max gas mileage" and who "drives more cautiously than most" respectively (Table 1). The distributions of driving style related factors are approximately similar for vehicle 2. Regarding the method of calculation, nearly half (50.2% and 43.0%) of vehicle 1 and vehicle 2 drivers calculated their fuel economy by own methods (e.g., simple math) respectively (Table 1).

[8] In particular, the efficiency gains obtained via multivariate framework presented in this study are a derivative of the correlation between the "residuals" or "unobserved factors". As discuss later, the residuals of the regressions for vehicle 1 and vehicle 2 gaps are significantly correlated and the residual dependencies in fact are stronger in the joint random-parameter framework.

However, a significant 35.6% and 36.4% of vehicle 1 and 2 drivers calculated their fuel economy either by miles and fuel purchase diary or odometer and fuel purchase diary (Table 1).

**(PUT FIGURE 1 ABOUT HERE)**

**(PUT FIGURE 2 ABOUT HERE)**

As expected, the majority of the two vehicles within the garage are gasoline. However, the percentages of vehicle 2 being gasoline vehicles is slightly smaller whereas the percentage of vehicle 2 being hybrid is greater than its vehicle 1 counterparts. This is expected given that hybrids are now penetrating the US market. Note that vehicle 2 is the newer vehicle (as discussed earlier). Of the first vehicle within a garage, 26.2% are manual compared to a lesser 19.6% manual vehicle 2 (Table 1). Intuitively, the share of CVT (continuously varying transmission) has increased for vehicle 2, which is the newer vehicle in the garage. From vehicle class, wheel drive, and manufacturer perspective, the My MPG two-vehicle garages sample seems to be reasonably representative (see descriptive statistics in Table 1).

## MODEL ESTIMATION RESULTS

All the fixed- and random-parameter regression models were derived from a systematic process to include most important variables (available in the data set) on the basis of statistical significance, specification parsimony, and intuition. Initially, a set of univariate regression models, separate for fuel economy gap of vehicle 1 and vehicle 2, were developed to conceptualize correlates of fuel economy gaps. Variables related to driving style, calculation method, vehicle type, transmission, fuel type, and manufacturer were included and those retained that were statistically significant at an at-least 80% significance level. The goodness of fit statistics are provided for the two final univariate models, one for vehicle 1 and other for vehicle 2, respectively (Table 2).

Next, a series of fixed-parameter bivariate seemingly unrelated regressions were estimated. Conceptually, the fixed parameter bivariate model allows for greater flexibility as it can account for potential correlations between fuel economy gaps of vehicle 1 and 2 within the same garage. Also, the bivariate model exhibited strong and statistically significant correlation in fuel economy gaps of the two vehicles, and which supports the motivation of using a multivariate modeling framework (Khattak et al., 1995; Yamamoto and Shankar, 2004a). As a result, more efficient parameter estimates (i.e., smaller standard errors) were obtained compared to univariate models and this is also reflected in the significant improvement in model goodness of fit of fixed parameter bivariate seemingly unrelated regression compared to the two univariate counterparts (Table 2). For instance, all the information criteria for the bivariate model are significantly smaller (after accounting for degrees of freedom) than the univariate counterparts, suggesting significantly better model fit of the joint model (Table 2).

**(PUT TABLE 2 ABOUT HERE)**

Finally, several random-parameter bivariate seemingly unrelated regression models were estimated (as discussed earlier). The joint random parameter model simultaneously accounts for the methodological concerns of cross-equation residual dependencies, and unobserved heterogeneity (i.e., unobserved factors varying systematically across the sample observations). As can be seen in Table 2, incorporating unobserved heterogeneity in the joint model estimation framework resulted in substantive improvements in model goodness-of-fit, i.e., all the information criteria scores are the lowest for random parameter bivariate seemingly unrelated regression model (Table 2). In addition, several variables that were statistically insignificant in fixed-parameter framework emerged statistically significant in the random-parameter counterpart. These findings provide compelling evidence that:

- There exists correlation among the degree of fuel economy gaps of vehicles within the same garage, and;
- The parameter estimates vary significantly across both garages and vehicles due to unobserved preference heterogeneity.

Table 3 presents the results of fixed- and random-parameter bivariate seemingly unrelated regression models. For the random parameter specification, all those explanatory variables ($\boldsymbol{X_i}$ or $\boldsymbol{B_i}$) are retained as random parameters if either of the following two criteria are met: 1) if both mean and standard deviation of the parameter density function are statistically significant at 95% confidence level, or 2) if the mean is statistically insignificant but the standard deviation of the parameter density function is statistically significant, and a model goodness-of-fit statistic (ICOMP, CAIC, or likelihood ratio test) suggesting that the model with $\boldsymbol{X_i}$ or $\boldsymbol{B_i}$ treated as random parameter (with zero mean and statistically significant standard deviation) is statistically better than the counterpart with $\boldsymbol{X_i}$ or $\boldsymbol{B_i}$ treated as fixed parameter[9].

A total of 14 variables (7 variables in the equation for vehicle 1 gap and 7 variables in the equation for vehicle 2 gap) were found to be normally distributed random parameters suggesting that their correlations vary significantly across the sampled garages (Table 3). For the random parameters, the distributional parameters, effects, and approximate ranges of the random parameters across the U.S. two-vehicle garages are provided in Table 4. Also, note that many variables that were statistically insignificant in the fixed-parameter joint model are now statistically significant in the random-parameter counterpart. All of these findings reveal the statistical supremacy of random parameter bivariate model, and the importance of addressing unobserved heterogeneity for establishing reliable and unbiased correlations (Mannering et al., 2016).

## DISCUSSION

We discuss findings based on the results of the random parameter bivariate model given its statistically superior fit (Table 2). The dependent variables for the models presented in Table 3 are the ratio of user-reported on-road MPG and EPA test cycle estimates. As both dependent variables and explanatory factors are untransformed (level-level regressions), the coefficients on explanatory factors can be interpreted as marginal effects. For continuous explanatory variables, a $\beta$ would translate to the unit increase (decrease) in the fuel economy gap for each one-unit increase in continuous explanatory variable, given all other variables remain constant. Likewise, if a dummy explanatory variable switch from zero to one, we would expect the gaps (of vehicle 1 or 2) to increase (decrease) by $\beta$ with other variables held at constant.

As was noted earlier in the section on bivariate distributions, an analysis of the observed fuel economy gaps of the two vehicles within the same garage revealed a positive Pearson correlation of 0.40.

[9] In most of the literature, a variable is treated as random if both mean and standard deviation of the density function are statistically significantly different from zero. However, the statistical significance of the mean of the density function is trivial if one can expect both positive and negative values of the parameter estimates. As we will see later, a variable with both statistically significant mean parameter (but smaller in magnitude) and large standard deviation will lead to a distribution with ~55% positive values and ~45% negative values (Milton et al., 2008; Li et al., 2017a). From an empirical stand-point, this scheme is similar to a statistically insignificant mean estimate but with a large (statistically significant) standard deviation – leading to half positive and half negative values. Such an estimation output simply implies substantial heterogeneity (not just in magnitudes but also in the direction of effects) in the effects of explanatory factors. As recognized recently by (Behnood and Mannering, 2017), this is a strong advantage of random parameters approach compared to fixed parameter counterpart. A mean estimate statistically indifferent from zero in fixed-parameter models will suggest that the specific explanatory variable is not statistically significant (or important). Contrarily, the random parameter framework will adequately capture the significantly varying effects in terms of direction. As we discuss later, this has important implications with regards to the interpretation of the effects of variables in the context of this study.

The models presented in Table 3 also quantifies the correlation in the error terms (unobserved factors) between the regressions for fuel economy gaps of vehicle 1 and 2. The results of joint random parameter model, shown in Table 3, reflect a correlation parameter of 0.519 (t-stat of 29.32). This suggests that the unobserved factors tend to jointly increase (decrease) the fuel economy gaps of vehicles within the same garage. This correlation may relate to factors such as income or education, which are not captured in the My MPG data, but may increase (decrease) the fuel economy gaps within a garage.

***Effects of Driving Style and Fuel Economy Calculation Method:***

Regarding driving style, compared to drivers that did not report their driving style, drivers who drive for maximum gas mileage, drive more cautiously than most, or drives with the traffic flow are likely to have greater on-road fuel economy relative to what EPA ratings suggest (Table 3). These findings suggest that drivers who are cautious or are "eco-drivers" can get more benefits as of higher on-road fuel economy (Kurani et al., 2013; Greene et al., 2017b; Greene et al., 2017 ). Contrarily, drivers who drive little faster than most, or who accelerate quickly and passes (volatile driving) are likely to have a gap below the average gaps (average gaps for vehicle 1 and vehicle 2 are 0.86 and 0.85 respectively – see Table 1). These findings hold both for vehicle 1 and 2 within the same garage and suggest that driving style matters.

The estimated models also quantify the effects of users' calculation methods used for estimation of on-road fuel economy. Both for vehicle 1 and 2, the four different methods considered in the joint random parameter regression models are, 1) Computer display – MPG noted from car's dashboard display, 2) miles & fuel purchase diary – MPG estimated by driving miles between fill-ups by gallons of fuel purchased, 3) Odometer & fuel purchase diary – MPG estimated by www.fueleconomy.gov by subtracting the odometer reading at the current fill up from the reading at the previous fill up and dividing by gallons purchased at the current fill up, and 4) Own method – user estimates his/her on-road MPG using own, unspecified method.

For vehicle 1, it is found that drivers who used own (and unspecified) methods for estimating their fuel economy are likely to have greater on-road fuel economy relative to what EPA ratings suggest ($\boldsymbol{\beta}$ = 0.0102). However, this finding should be interpreted with caution as own unspecified methods used by drivers for calculation of fuel economy may be inaccurate and/or drivers may over-estimate their actual on-road fuel economy. This variable was found statistically insignificant in vehicle 2 equation.

Contrarily, drivers who used computer display, miles & fuel purchase diary, or odometer & fuel purchase diary are likely to experience gaps below the average gaps. However, for vehicle 1, the effects of odometer & fuel purchase diary on fuel economy gap exhibit significant heterogeneity with a mean of -0.0285 and standard deviation of 0.0435 (Table 3). For 74.43% of the garages, calculations based on odometer & fuel purchase diary are associated with gap below the average gaps (i.e., users getting significantly lower on-road fuel economy relative to EPA ratings). Whereas, for 25.57% of the garages, this method of calculation is associated with greater on-road fuel economy (relative to EPA ratings) (Table 4). This heterogeneity is also reflected in the largely varying parameters estimates with lower and upper range of (approximately) -0.1153 and 0.0583 respectively (Table 4). Also, note that the parameter estimates for miles & fuel purchase diary and odometer & fuel purchase diary are similar. This is expected as both methods are based on same data, i.e., fuel purchases and miles driven between successive fuel purchases (Greene et al., 2017b).

**(PUT TABLE 3 ABOUT HERE)**

***Effects of Fuel Type, Transmission, and Wheel Drive:***

Compared to diesel vehicles, hybrids on average tend to have a greater gap ($\boldsymbol{\beta}$ = -0.0934 and -0.1000 for vehicle 1 and 2 respectively). Turning to gasoline vehicles, the parameter estimates (both for vehicle 1 and 2) exhibit significant heterogeneity. For 59.81% of vehicle 1, gasoline fuel type is associated with better on-road fuel economy relative to EPA ratings, however, for a significant 40.19% of vehicle 1 gasoline fuel type is associated with greater gap (see distributions in Table 4). Likewise, for vehicle 2, 62.77% of gasoline fuel type vehicles exhibited greater gap whereas 37.23% exhibited better on-road fuel economy relative to EPA test cycle estimates Table 4. This significant heterogeneity suggests that fuel type all alone cannot be attributed to lower or higher gap and that several other unobserved factors may contribute to the gap. Note that this variable is statistically insignificant in the fixed parameter joint model (Table 3) and which reveals the statistical superiority of the random parameter counterpart[10].

Coming to transmission type, both for vehicle 1 and 2, vehicles with CVT transmission and automatic transmission have on-average higher gaps (Table 3). These findings are in agreement with the literature (Schneider et al., 1982). However, the effects of an automatic transmission with gears < 7 speed on fuel economy gap exhibit significant heterogeneity with parameter estimates ranging between -0.0937 and 0.0163, i.e., random parameter. Two-wheel drive (either front or rear) tend to have greater on-road fuel economy relative to EPA ratings, whereas all-wheel drive vehicles tend to have greater gaps. Again, the variable rear-wheel drive in vehicle 2 equation is found normally distributed random parameter, with positive and negative associations for 56.87% and 43.13% of the garages respectively. All these findings provide compelling evidence of unobserved heterogeneity in the sample and ignoring which can result in highly biased parameter estimates.

***Effects of Vehicle Class and Manufacturer:***

The parameter estimates for larger body types are generally negative suggesting greater gaps for larger vehicles. Contrarily, for two-seaters, the on-road fuel economy on-overage is greater than the EPA test cycle estimates (Table 3). These findings are in agreement with the literature (Greene et al., 2017b; Greene et al., 2017 ). Nonetheless, subcompact and two-seater variables are found to be normally distributed random parameters in the vehicle 1 equation. The range of both parameter estimates is large, and which shows that the effects of these variables on fuel economy gap vary significantly (Table 4).

The car manufacturer is negatively associated with the response outcomes for Ford, GM, Toyota, Nissan, Chrysler, and Hyundai, suggesting larger gap than average for these manufacturers. On the other hand, Daimler and VW on-average experience higher on-road fuel economy than the EPA test cycle estimates. Nonetheless, several of the manufacturer effects are not unidirectional with the effects of Chrysler, Ford, GM, and VW varying significantly across the sampled garages (Table 4). The random parameters for several manufacturers are intuitive as attributing greater or smaller gap solely to vehicle manufacturing company (irrespective of garage characteristics, travel needs, and driving styles etc.) may be unrealistic.

## LIMITATIONS

[10] Simple univariate fixed-parameter regressions for vehicle 1 and vehicle 2 gaps with gasoline fuel type as only explanatory factor resulted in statistically significant parameter estimates of 0.0530 and 0.0734 respectively. This suggests that the association between gasoline fuel type and fuel economy gaps of the two vehicles are constant and positive across the sampled households. However, after accounting for potential correlation and unobserved heterogeneity, the mean parameter estimates (and not the standard deviations) in the random-parameter joint model become statistically insignificant. This provides compelling evidence that the positive statistically significant parameter estimates in univariate models were an outgrowth of unobserved factors, which are now tracked as unobserved heterogeneity in the joint random-parameter model.

A disadvantage of the sample is the self-selected and non-random nature of the fueleconomy.gov's My MPG database. For a self-selected sample, an argument can be made that users who participated in fueleconomy.gov's My MPG project have a higher tendency to exceed their EPA test cycle estimates than those who did not participate in the My MPG project. That is, important unobservable factors may influence whether a garage appears in the My MPG sample and/or such unobservable factors may be correlated with the covariates included in the regressions for the non-random sample. As discussed in detail earlier, this may be viewed as an omitted variables specification problem (Heckman, 1977; Heckman, 2010). A recent study by Greene et al. (2017) found that self-selection bias may be present in the My MPG sample; however, it did not appear to strongly influence the parameter estimates (Greene et al., 2017b). To correct for the self-selection bias problem, several estimators such as the Heckman's model and instrumental variables techniques can be considered. Unfortunately, nearly all estimators for correcting self-selection require that we have at least some data on a random sample from the relevant population (i.e., garages who did not participate in the My MPG project). Without this data, we cannot test or correct for sample selection. Given these constraints, a rigorous remedy is to consider heterogeneity modeling framework to account for the omitted variables bias. In the models presented, the omitted explanatory factors become a portion of the unobserved heterogeneity, and which has the potential to mitigate adverse impacts of omitting important explanatory factors. However, as noted in (Mannering et al., 2016), we acknowledge that the parameter estimates obtained from the joint heterogeneity based models may not track the unobserved heterogeneity (due to omitted variables) as good as if we could have included the important omitted variables in the model specification. Nonetheless, keeping in view the data constraints and which precludes estimation of common self-selection correction models, the methodology adopted in this study is expected to be a competitive methodological alternative.

**(PUT TABLE 4 ABOUT HERE)**

## CONCLUSIONS/PRACTICAL IMPLICATIONS

The main objectives of this study were to investigate 1) the degree of interrelationship between fuel economy gaps of vehicles owned by the same garage, 2) how the fuel economy gaps vary within and across different garages, and 3) the differential effects of several factors such as vehicle characteristics and driving behavior on fuel economy gaps of vehicles within the same garage. To achieve this, an in-depth statistical analysis is conducted on users' self-reported fuel economy data vis-a-vis EPA fuel economy ratings for more than 7000 garages who voluntarily submitted data to the "My MPG" section of the government website www.fueleconomy.gov.

The empirical analysis, nonetheless, is complicated owing to the presence of important methodological concerns including potential interrelationships between fuel economy gaps of vehicles in the same garage and unobserved taste or preference heterogeneity. Thus, the present study empirically contributes by addressing unobserved heterogeneity in a multivariate empirical setting. Specifically, the bivariate seemingly unrelated fixed parameter and random parameter modeling approach is presented that accounts for the presence of correlation within a garage and the possibility to allow parameter estimates to vary across vehicles and garages. By simultaneously modeling the ratio of user-reported fuel economy and EPA test cycle estimates for the two vehicles, a fundamental understanding is sought regarding the differential effects of several factors on fuel economy gaps of vehicles within the same garage. To authors' knowledge, such empirical methods are not used in the context under discussion. Several findings emerged from the analysis:

- For the two vehicles, the average gap is 13.1% and 14.8% for vehicle 1 and 2 respectively. That is, for the two vehicles, user-reported on-road fuel economy on average is 86.9% and 85.2% of the official EPA test cycle estimates. With EPA test cycle ratings tending to over-estimate the actual on-road fuel economy, a significant variation is observed in the fuel economy gaps for the two vehicles across the sampled garages. That is, for vehicle 1, the on-road fuel economy was observed to be as low as 36% and as high as 137% of the official EPA ratings. Similar variation was also observed for vehicle 2.
- The analysis also revealed that the fuel economy gaps are increasing for the newer model year vehicles, with the gap increasing since 2005 with highest gaps during 2009-2014. For example, compared to average gaps of 13.1% and 14.8% for the two vehicles, the highest gaps of 18.7% and 19.3% for vehicle 1 and 2 occurred for vehicles with model years between 2009-2014.
- It is found that the fuel economy gaps of two vehicles within the same garage are mildly positively correlated with a Pearson correlation of 0.40. The density of the bivariate correlation is higher in the middle of the distribution (fuel economy gaps between 0.75 and 0.9) with smaller density elsewhere. From a policy perspective, this implies lack of compelling agreement between the fuel economy gaps which could weaken consumers' confidence in making relative comparisons among vehicles.
- Using a joint random parameters bivariate seemingly unrelated regression equations (RP-SURE) framework, to account for unobserved heterogeneity across garages, and for the interdependencies of fuel economy gaps within a garage, the differential effects of a wide variety of factors on fuel economy gaps of two vehicles are quantified including driving style, fuel economy calculation method, fuel type, vehicle transmission, wheel drive, vehicle class, and vehicle manufacturer.
- Drivers who drive for maximum gas mileage, drive more cautiously than most, or drives with the traffic flow are likely to have greater on-road fuel economy relative to what EPA ratings suggest. Contrarily, drivers who drive little faster than most, or who accelerate quickly and passes (volatile driving) are likely to have larger gaps. These findings suggest that drivers who are cautious or are "eco-drivers" can get more benefits in form of higher on-road fuel economy. Compared to diesel vehicles, hybrids on average tend to have a greater gap. The correlations between gasoline vehicles and fuel economy gaps exhibited significant heterogeneity not just in the magnitudes of the effects but also direction of effects. This suggests that higher (or lower) gaps may not be solely attributed to fuel type. Regarding transmission type, vehicles with CVT transmission and automatic transmission have on average greater gaps. Finally, larger body type vehicles are observed to exhibit greater gaps, whereas, for two-seaters, the on-road fuel economy on-overage is greater than the EPA test cycle estimates. Overall, the differential effects of several explanatory factors on fuel economy gaps of the two vehicles exhibited substantial heterogeneity across the sampled garages and are discussed in detail. The model results showed that RP-SURE model provides significantly superior fit compared to a series of fixed and random parameter univariate linear regressions as well as a bivariate model with fixed parameters. Keeping in view the compelling statistical evidence, the results clearly emphasize that it is crucial to account for both within garage correlation and the heterogeneity in parameter estimates in order to reach unbiased and reliable inferences and to avoid under- or overstating the effects of specific explanatory variables.

The results of this study have practical implications. First, the finding that fuel economy gaps are increasing since 2005 with significant gap during 2009-2014 suggests that the expected benefits of greenhouse gas regulations and fuel economy cannot be fully realized. Second, a mild degree of correlation is found between the fuel economy gaps of two vehicles within a garage. Analysis of the deviations for pairs of vehicles in the same garage suggests that the EPA test cycle ratings are useful but not as useful as one would like them to be. The strength of association overall is weak, creating uncertainty even in relative comparisons among vehicles. From a policy perspective, this finding raises important questions about the greater usefulness of official EPA ratings for vehicle comparison purposes. Related to the effects of explanatory factors on fuel economy gaps, the findings suggest that driving style matters. For instance, vehicles within the garage who are driven by more volatile drivers (e.g., accelerates quickly and pass, or drives little faster than most) are likely to experience greater gaps. This points out to incorporating individual level driving volatility indices in driving behavior monitoring and feedback devices as reflected in efforts by the authors (Wang et al., 2015; Liu et al., 2016a; Liu and Khattak, 2016). For example, mechanisms to monitor driving behavior in real-time can be developed and real-time warnings can be given to drivers if the driver is observed to be volatile, potentially reducing fuel consumption.

## ACKNOWLEDGEMENT

This paper is based upon work supported by the U.S. Department of Energy and Oak Ridge National Laboratory. The support of University of Tennessee's Transportation Engineering and Science Program and Initiative for Sustainable Mobility, a campus-wide organized research unit, is gratefully appreciated. Any opinions or recommendations expressed in this paper are those of the authors. The authors would also like to recognize the contributions of Ms. Alexandra Boggs and Ms. Megan Lamon in proof-reading several versions of the manuscript. Finally, the authors acknowledge the helpful comments of two anonymous reviewers on an earlier version of the paper.

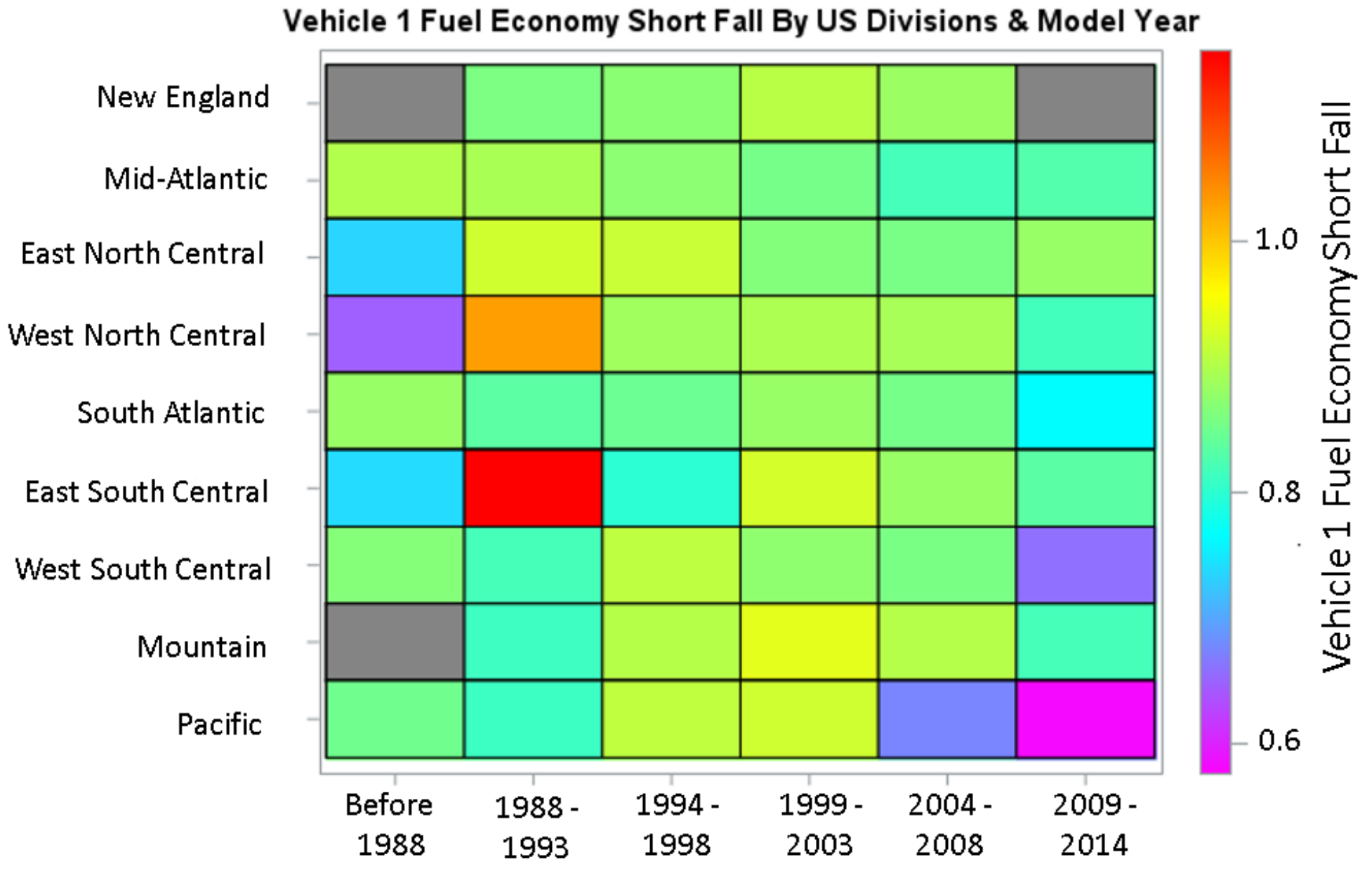

**69.**

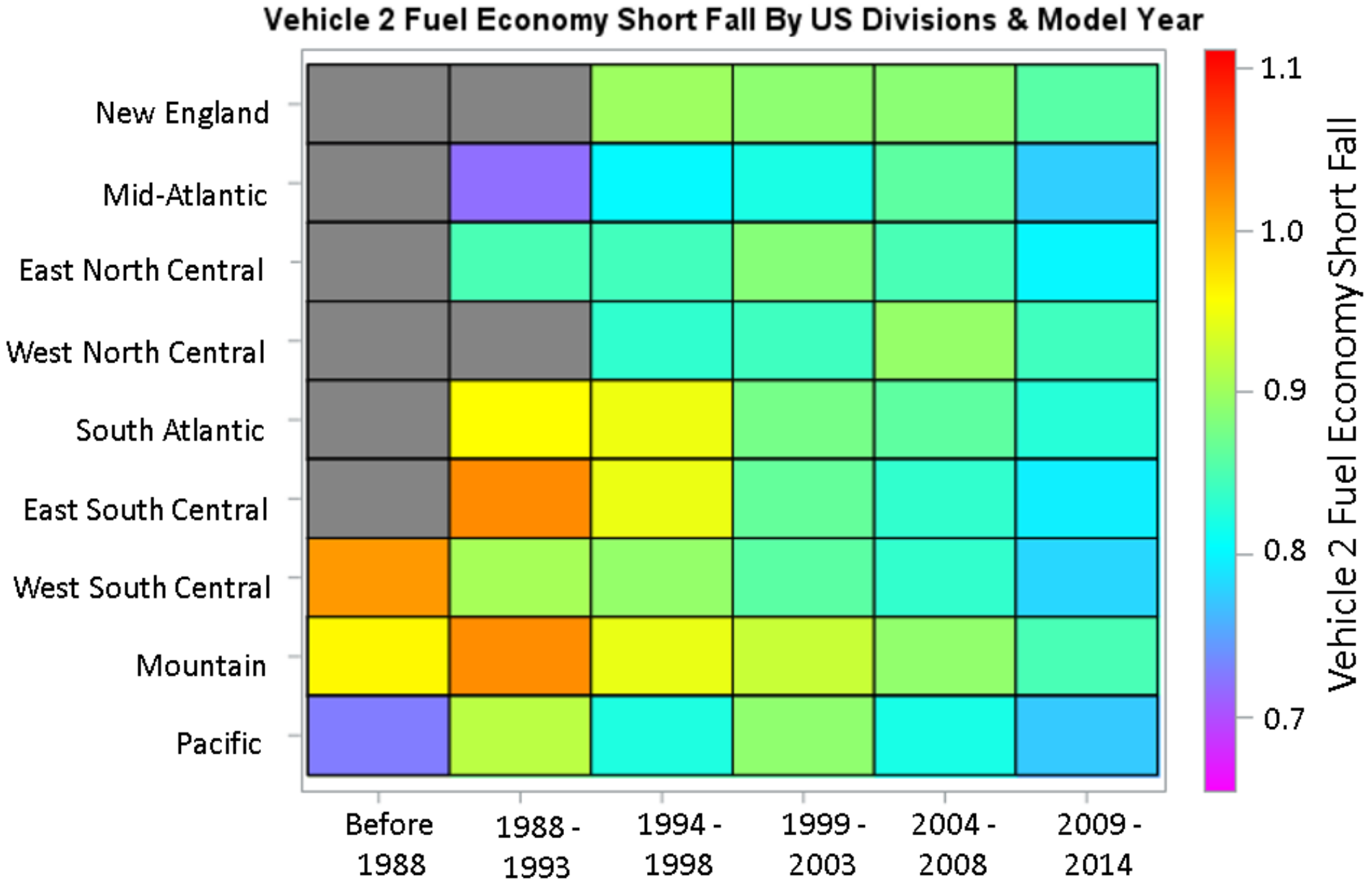

**FIGURE 1: Fuel Economy Gaps by US Division and Model Year**

(Notes: Top panel - Calendar heat map for vehicle 1; Bottom panel - Calendar heat map for vehicle 2; Grey cells indicate no data available; For definitions of nine US divisions, see https://en.wikipedia.org/wiki/List_of_regions_of_the_United_States).

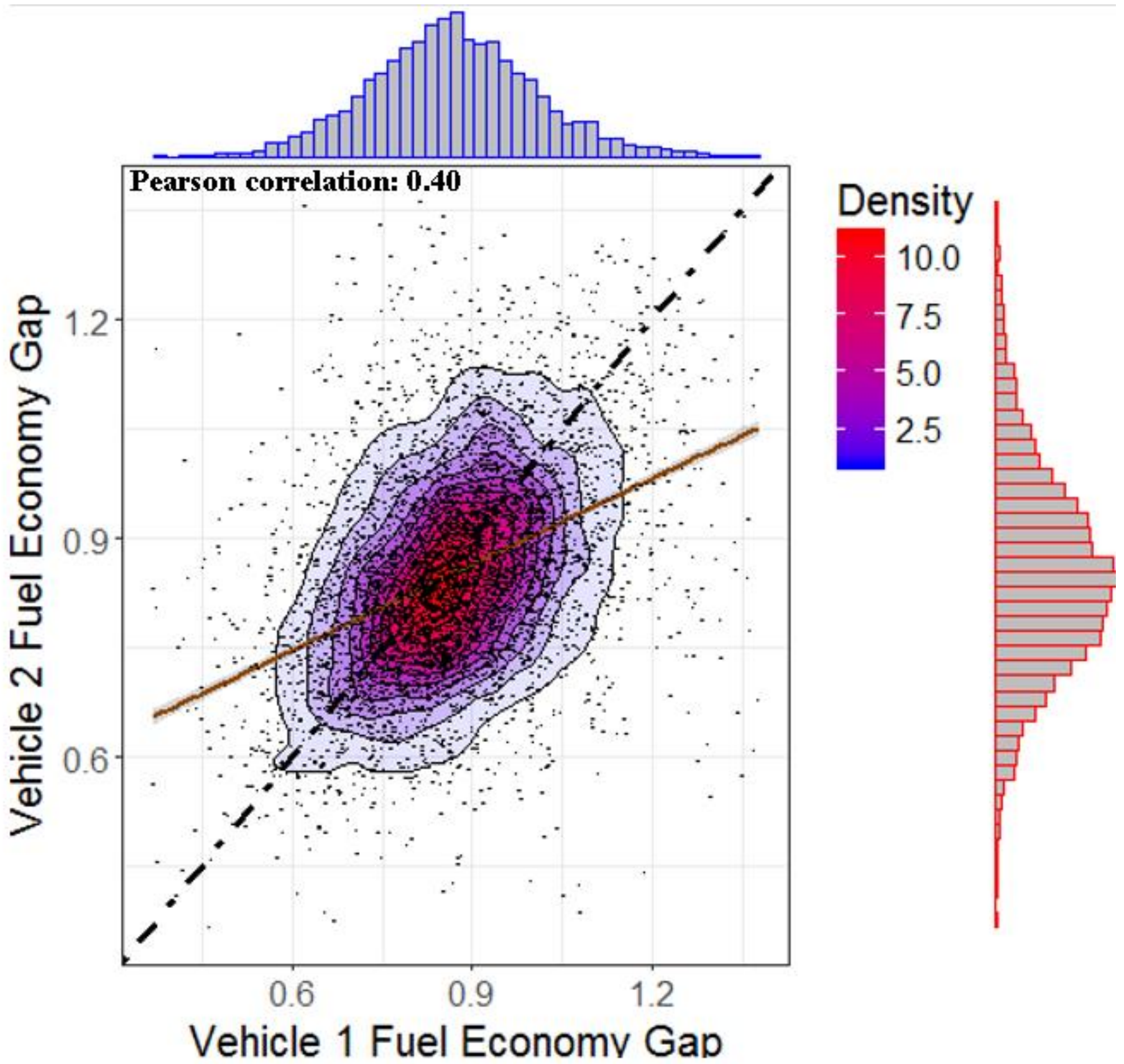

**FIGURE 2: Bivariate Distribution and Marginal Distributions of Fuel Economy Gaps**
(Notes: Black dashed line represents y=x line; Brown solid line represents linear correlation between the fuel economy gaps of the two vehicles)

## TABLE 1: Descriptive Statistics of Key Variables

| Variable | Vehicle 1 | | | Vehicle 2 | | |
|---|---|---|---|---|---|---|
| | Mean | SD | Min/Max | Mean | SD | Min/Max |
| **Untrimmed Sample (N = 7244)** | | | | | | |
| Average My MPG | 24.71 | 8.54 | 2.86/96.093 | 26.21 | 9.05 | 3/100.11 |
| EPA ratings | 28.41 | 8.89 | 11.95/76.27 | 30.96 | 10.70 | 12.47/76.27 |
| Fuel economy gap (%) | 0.87 | 0.17 | 0.12/3.55 | 0.86 | 0.17 | 0.16/3.92 |
| **Trimmed Sample (N = 7126)** | | | | | | |
| Average My MPG | 24.63 | 8.23 | 6.61/83 | 26.09 | 8.76 | 8.06/91.30 |
| EPA ratings | 28.44 | 8.91 | 11.95/76.27 | 30.99 | 10.71 | 13.05/76.27 |
| Fuel economy gap (%) | 0.86 | 0.14 | 0.36/1.37 | 0.85 | 0.14 | 0.36/1.36 |
| **Model Year** | | | | | | |
| 1984-1988 | 0.048 | 0.21 | 0/1 | 0.004 | 0.06 | 0/1 |
| 1989-1993 | 0.11 | 0.31 | 0/1 | 0.02 | 0.15 | 0/1 |
| 1994-1998 | 0.22 | 0.42 | 0/1 | 0.08 | 0.26 | 0/1 |
| 1999-2003 | 0.32 | 0.47 | 0/1 | 0.22 | 0.41 | 0/1 |
| 2004-2008 | 0.25 | 0.43 | 0/1 | 0.47 | 0.50 | 0/1 |
| 2009-2014 | 0.05 | 0.21 | 0/1 | 0.21 | 0.41 | 0/1 |
| **Driving Style** | | | | | | |
| Drives for max gas mileage | 0.007 | 0.083 | 0/1 | 0.008 | 0.088 | 0/1 |
| Drives more cautiously than most | 0.054 | 0.226 | 0/1 | 0.055 | 0.228 | 0/1 |
| Drives with the flow | 0.083 | 0.276 | 0/1 | 0.093 | 0.290 | 0/1 |
| Drives a little faster than most | 0.117 | 0.322 | 0/1 | 0.124 | 0.330 | 0/1 |
| Accelerates quickly and passes | 0.065 | 0.247 | 0/1 | 0.064 | 0.244 | 0/1 |
| Not reported | 0.674 | 0.131 | 0/1 | 0.637 | 0.091 | 0/1 |
| **Calculation Method** | | | | | | |
| Computer display | 0.008 | 0.089 | 0/1 | 0.009 | 0.093 | 0/1 |
| Miles & fuel purchase diary | 0.160 | 0.366 | 0/1 | 0.161 | 0.368 | 0/1 |
| Odometer & fuel purchase diary | 0.196 | 0.397 | 0/1 | 0.203 | 0.402 | 0/1 |
| Own method | 0.502 | 0.500 | 0/1 | 0.430 | 0.495 | 0/1 |
| **Fuel Type** | | | | | | |
| Diesel | 0.021 | 0.144 | 0/1 | 0.028 | 0.164 | 0/1 |
| Gasoline | 0.943 | 0.232 | 0/1 | 0.895 | 0.306 | 0/1 |
| Hybrid | 0.036 | 0.185 | 0/1 | 0.077 | 0.267 | 0/1 |
| **Transmission** | | | | | | |
| Manual | 0.262 | 0.440 | 0/1 | 0.196 | 0.397 | 0/1 |
| CVT | 0.042 | 0.200 | 0/1 | 0.101 | 0.301 | 0/1 |
| Automatic (Gears ≥ 7 speed) | 0.003 | 0.058 | 0/1 | 0.010 | 0.099 | 0/1 |
| Automatic (Gears < 7 speed) | 0.693 | 0.461 | 0/1 | 0.692 | 0.462 | 0/1 |
| **Wheel drive** | | | | | | |
| Four-wheel drive | 0.228 | 0.121 | 0/1 | 0.213 | 0.321 | 0/1 |
| All-wheel drive | 0.005 | 0.072 | 0/1 | 0.024 | 0.152 | 0/1 |
| Front-wheel drive | 0.560 | 0.496 | 0/1 | 0.623 | 0.485 | 0/1 |
| Rear-wheel drive | 0.206 | 0.405 | 0/1 | 0.140 | 0.347 | 0/1 |

Notes: SD is standard deviation; N is sample size.

**TABLE 1:** ***(Continued).***

| Variable | Vehicle 1 | | | Vehicle 2 | | |
|---|---|---|---|---|---|---|
| | **Mean** | **SD** | **Min/Max** | **Mean** | **SD** | **Min/Max** |
| **Vehicle Class** | | | | | | |
| Compact | 0.211 | 0.408 | 0/1 | 0.185 | 0.388 | 0/1 |
| Large | 0.053 | 0.224 | 0/1 | 0.054 | 0.226 | 0/1 |
| Midsize | 0.180 | 0.384 | 0/1 | 0.208 | 0.406 | 0/1 |
| Minivan | 0.034 | 0.182 | 0/1 | 0.056 | 0.230 | 0/1 |
| Pickup | 0.122 | 0.327 | 0/1 | 0.083 | 0.276 | 0/1 |
| SUV | 0.151 | 0.358 | 0/1 | 0.222 | 0.416 | 0/1 |
| Special purpose | 0.059 | 0.236 | 0/1 | 0.020 | 0.141 | 0/1 |
| Subcompact | 0.122 | 0.328 | 0/1 | 0.103 | 0.304 | 0/1 |
| Two-seaters | 0.029 | 0.169 | 0/1 | 0.018 | 0.133 | 0/1 |
| Vans | 0.010 | 0.100 | 0/1 | 0.004 | 0.065 | 0/1 |
| Wagons | 0.028 | 0.166 | 0/1 | 0.046 | 0.210 | 0/1 |
| **Manufacturer** | | | | | | |
| Ford | 0.138 | 0.345 | 0/1 | 0.127 | 0.333 | 0/1 |
| GM | 0.208 | 0.406 | 0/1 | 0.169 | 0.374 | 0/1 |
| Toyota | 0.136 | 0.343 | 0/1 | 0.170 | 0.375 | 0/1 |
| Nissan | 0.052 | 0.221 | 0/1 | 0.050 | 0.219 | 0/1 |
| Chrysler | 0.109 | 0.311 | 0/1 | 0.093 | 0.290 | 0/1 |
| Honda | 0.131 | 0.338 | 0/1 | 0.146 | 0.353 | 0/1 |
| BMW | 0.023 | 0.151 | 0/1 | 0.027 | 0.162 | 0/1 |
| VW | 0.050 | 0.218 | 0/1 | 0.052 | 0.222 | 0/1 |
| Hyundai | 0.024 | 0.154 | 0/1 | 0.048 | 0.213 | 0/1 |
| Daimler | 0.017 | 0.129 | 0/1 | 0.013 | 0.115 | 0/1 |
| Other | 0.111 | 0.181 | 0/1 | 0.107 | 0.143 | 0/1 |
| Displacement (liter) | 2.977 | 1.171 | 1/8.3 | 2.882 | 1.125 | 1/8.3 |
| Turbo | 0.061 | 0.240 | 0/1 | 0.075 | 0.264 | 0/1 |

Notes: SD is standard deviation.

**TABLE 2: Models Goodness of Fit**

| Models | N | LL | DF | Bozdogan's CAIC | Bozdogan's ICOMP | AIC | BIC |
|---|---|---|---|---|---|---|---|
| Fixed Parameter Univariate - Vehicle 1 | 7126 | 4109.059 | 30 | -8193.364 | -7931.873 | -8160.118 | -7960.873 |
| Fixed Parameter Univariate - Vehicle 2 | 7126 | 4470.486 | 34 | -8910.978 | -8615.211 | -8874.971 | -8648.211 |
| *Summation* | | *8579.545* | *64* | *-17104.342* | *-16547.084* | *-17035.089* | *-16609.084* |
| Fixed Parameter Bivariate Seemingly Unrelated Regression | 7126 | 9153.8972 | 67 | -18252.88 | -17695.76 | -18183.79 | -17757.76 |
| Random Parameter Bivariate Seemingly Unrelated Regression | 7126 | 9239.2743 | 81 | -18291.05 | -17856.64 | -18352.55 | -17919.64 |

**Notes: N** is sample size; LL is log-likelihood at convergence; DF is degrees of freedom; CAIC Consistent Akaike's Information Criteria; ICOMP is Bozdogan's index of information complexity; AIC is Akaike's Information Criteria; and BIC is Schwartz Bayesian Information Criteria.

**TABLE 3: Model Estimation Results - Random parameters bivariate seemingly unrelated regression equations and Fixed parameters bivariate seemingly unrelated regression equations.**

| Variable | Fixed Parameters Bivariate Model | | | | Random Parameters Bivariate Model | | | |
|---|---|---|---|---|---|---|---|---|
| | Vehicle 1 | | Vehicle 2 | | Vehicle 1 | | Vehicle 2 | |
| | ***β*** | **t-stat** | ***β*** | **t-stat** | ***β*** | **t-stat** | ***β*** | **t-stat** |
| Constant | 0.8796 | 56.45 | 0.9212 | 67.71 | 0.8774 | 59.54 | 0.9229 | 70.15 |
| Cross-Equation Correlation[b] | 0.4112941 (34.359) | | | | 0.519 (29.32) | | | |
| ***Driving Style*** | | | | | | | | |
| Drives for max gas mileage | 0.0686 | 3.79 | 0.0487 | 2.89 | 0.0692 | 3.86 | 0.0477 | 3.14 |
| Drives more cautiously than most | 0.0565 | 8.31 | 0.0481 | 7.5 | 0.0555 | 8.31 | 0.0597 | 7.71 |
| Drives with the flow | 0.0154 | 2.74 | 0.0104 | 2.05 | 0.0149 | 2.69 | 0.0108 | 2.17 |
| Drives a little faster than most | -0.0162 | -3.36 | -0.0162 | -3.58 | -0.0154 | -3.22 | -0.0163 | -3.69 |
| Accelerates quickly and passes | -0.0244 | -3.89 | -0.0343 | -5.71 | -0.0240 | -3.87 | -0.0347 | -5.84 |
| Not reported (Base) | --- | --- | --- | --- | --- | --- | --- | --- |
| ***Calculation Method*** | | | | | | | | |
| Computer display | --- | --- | -0.0442 | -2.9 | --- | --- | -0.0449 | -3.02 |
| Miles & fuel purchase diary | -0.0372 | -6.47 | -0.0487 | -11.57 | -0.0375 | -6.7 | -0.0469 | -11.47 |
| Odometer & fuel purchase diary[a] | -0.0284 | -5.12 | -0.0421 | -10.81 | -0.0285 | -5.2 | -0.0433 | -11.51 |
| *standard deviation*[c] | --- | --- | --- | --- | *0.0435* | *5.16* | --- | --- |
| Own method | 0.0106 | 2.27 | --- | --- | 0.0102 | 2.23 | --- | --- |
| ***Fuel Type*** | | | | | | | | |
| Diesel (Base) | --- | --- | --- | --- | --- | --- | --- | --- |
| Gasoline[a] | 0.0105 | 0.87 | -0.0181 | -1.66 | 0.0129 | 1.17 | -0.0203 | -1.96 |
| *standard deviation*[c] | --- | --- | --- | --- | *0.0521* | *4.79* | *0.0624* | *12.01* |
| Hybrid | -0.0971 | -5.49 | -0.0986 | -7.11 | -0.0934 | -5.75 | -0.1000 | -7.59 |
| ***Transmission*** | | | | | | | | |
| CVT | -0.0606 | -5 | -0.0846 | -9.97 | -0.0625 | -5.46 | -0.0856 | -10.48 |
| Automatic (Gears ≥ 7 speed) | --- | --- | -0.0503 | -3.37 | --- | --- | -0.0474 | -3.36 |
| Automatic (Gears < 7 speed) [a] | -0.0386 | -9.89 | -0.0470 | -11.71 | -0.0387 | -9.99 | -0.0485 | -12.09 |
| *standard deviation*[c] | --- | --- | --- | --- | *0.0276* | *1.94* | --- | --- |
| ***Wheel Drive*** | | | | | | | | |
| All-wheel drive | --- | --- | -0.0241 | -2.39 | --- | --- | -0.0232 | -2.39 |
| Front-wheel drive | 0.0194 | 4.01 | 0.0096 | 1.77 | 0.0197 | 4.08 | 0.0082 | 1.52 |
| Rear-wheel drive[a] | --- | --- | 0.0088 | 1.58 | --- | --- | 0.0080 | 1.37 |
| *standard deviation*[c] | --- | --- | --- | --- | --- | --- | *0.0461* | *5.44* |
| **Vehicle Class** | | | | | | | | |
| Minivan | -0.0521 | -6.05 | -0.0597 | -8.74 | -0.0512 | -5.89 | -0.0591 | -8.64 |
| Pickup | -0.0486 | -7.69 | -0.0497 | -7.33 | -0.0477 | -7.56 | -0.0505 | -7.12 |
| SUV | -0.0226 | -4.21 | -0.0236 | -4.92 | -0.0216 | -4.05 | -0.0234 | -4.86 |
| Special purpose | -0.0376 | -5.14 | -0.0246 | -2.28 | -0.0372 | -5.03 | -0.0242 | -2.17 |
| Subcompact[a] | -0.0097 | -1.93 | -0.0170 | -3.22 | -0.0100 | -1.93 | -0.0165 | -3.14 |
| *standard deviation*[c] | --- | --- | --- | --- | *0.0506* | *5.13* | --- | --- |
| Two-seaters[a] | 0.0143 | 1.48 | 0.0205 | 1.75 | 0.0141 | 1.48 | 0.0186 | 1.55 |
| *standard deviation*[c] | --- | --- | --- | --- | *0.0285* | *2.31* | *0.0574* | *2.96* |

**Notes:** ***β*** is parameter estimate; (---) means not applicable; (a) indicates random parameters; (b) Enclosed in parenthesis are t-statistics; (c) Standard deviation of parameter density function.

**TABLE 3: (*Continued*).**

| Variable | Fixed Parameters Bivariate Model | | | | Random Parameters Bivariate Model | | | |
|---|---|---|---|---|---|---|---|---|
| | Vehicle 1 | | Vehicle 2 | | Vehicle 1 | | Vehicle 2 | |
| | ***β*** | **t-stat** | ***β*** | **t-stat** | ***β*** | **t-stat** | ***β*** | **t-stat** |
| Vans | -0.0550 | -3.55 | -0.0806 | -3.61 | -0.0573 | -3.68 | -0.0791 | -3.26 |
| Wagons | -0.0311 | -3.34 | -0.0218 | -3.04 | -0.0300 | -3.3 | -0.0218 | -3.09 |
| **Manufacturer** | | | | | | | | |
| Ford[a] | -0.0173 | -3.75 | -0.0187 | -3.17 | -0.0167 | -3.65 | -0.0189 | -3.21 |
| *standard deviation*[c] | --- | --- | --- | --- | --- | --- | *0.0432* | *4.83* |
| GM[a] | --- | --- | -0.0115 | -2.03 | --- | --- | -0.0113 | -1.96 |
| *standard deviation*[c] | --- | --- | --- | --- | --- | --- | *0.0594* | *8.85* |
| Toyota | --- | --- | -0.0246 | -4.38 | --- | --- | -0.0245 | -4.54 |
| Nissan | -0.0137 | -1.98 | -0.0318 | -4.13 | -0.0142 | -2.06 | -0.0315 | -4.21 |
| Chrysler[a] | -0.0054 | -1.04 | -0.0211 | -3.24 | -0.0054 | -0.98 | -0.0211 | -3.1 |
| *standard deviation*[c] | --- | --- | --- | --- | *0.0527* | *5.22* | *0.0676* | *8.48* |
| Honda | --- | --- | -0.0175 | -3.1 | --- | --- | -0.0167 | -3.09 |
| BMW | 0.0641 | 6.1 | --- | --- | 0.0652 | 6.19 | --- | --- |
| VW[a] | --- | --- | 0.0115 | 1.31 | --- | --- | 0.0106 | 1.2 |
| *standard deviation*[c] | --- | --- | --- | --- | --- | --- | *0.0513* | *4.90* |
| Hyundai | -0.0214 | -2.16 | -0.0398 | -5.11 | -0.0210 | -2.16 | -0.0383 | -5.14 |
| Daimler | 0.0369 | 2.96 | 0.0132 | 6.49 | 0.0384 | 3.14 | --- | --- |
| ***Displacement (liter)*** | 0.0071 | 3.86 | 0.0132 | 6.49 | 0.0070 | 3.8 | 0.0140 | 6.67 |
| ***Turbo***[a] | 0.0321 | 4.37 | --- | --- | 0.0336 | 4.64 | --- | --- |
| *standard deviation*[c] | --- | --- | --- | --- | 0.0349 | 2.08 | --- | --- |

**Notes:** ***β*** is parameter estimate; (---) means not applicable; (a) indicates random parameters; (c) Standard deviation of parameter density function.

**TABLE 4: Distributional parameters, effects, and approximate ranges of the random parameters across the U.S. two-vehicle garages for Random parameters bivariate seemingly unrelated regression equations**

| Variables | Distributional parameters for random held variables | | Approximate range of random parameters | | Percent of observations with individual for Random Parameters | |
|---|---|---|---|---|---|---|
| | $\mu$ | $\sigma$ | Lower | Upper | Above 0 | Below 0 |
| **Vehicle 1** | | | | | | |
| Odometer & fuel purchase diary | -0.02851 | 0.0434 | -0.1153 | 0.0583 | 25.57% | 74.43% |
| Gasoline | 0.01294 | 0.0521 | -0.0913 | 0.1172 | 59.81% | 40.19% |
| Automatic (Gears < 7 speed) | -0.03870 | 0.0275 | -0.0937 | 0.0163 | 7.97% | 92.03% |
| Subcompact | -0.03722 | 0.0505 | -0.1382 | 0.0638 | 23.06% | 76.94% |
| Two-seaters | 0.01411 | 0.0284 | -0.0427 | 0.0709 | 69.04% | 30.96% |
| Chrysler | -0.00537 | 0.0526 | -0.1106 | 0.0998 | 45.94% | 54.06% |
| Turbo | 0.03358 | 0.0349 | -0.0362 | 0.1034 | 16.80% | 83.20% |
| **Vehicle 2** | | | | | | |
| Gasoline | -0.02033 | 0.0624 | -0.1451 | 0.1044 | 37.23% | 62.77% |
| Rear-wheel drive | 0.00798 | 0.0461 | -0.0842 | 0.1002 | 56.87% | 43.13% |
| Two-seaters | 0.01856 | 0.0574 | -0.0963 | 0.1334 | 37.32% | 62.68% |
| Ford | -0.01891 | 0.0432 | -0.1053 | 0.0675 | 33.08% | 66.92% |
| GM | -0.01135 | 0.0594 | -0.1301 | 0.1074 | 42.43% | 57.57% |
| Chrysler | -0.02114 | 0.0676 | -0.1564 | 0.1141 | 37.72% | 62.28% |
| VW | 0.01062 | 0.0513 | -0.0920 | 0.1133 | 41.80% | 58.20% |

Notes: $\mu$ and $\sigma$ are the mean and standard deviation of the random parameters.